\documentclass[aps,prb,twocolumn,floatfix]{revtex4}
\usepackage{graphicx}
\begin{document}
\title{Resonating Valence Bond Theory of Superconductivity
for Dopant Carriers: Application to the Cobaltates}
\author{Alvaro Ferraz}
\affiliation{International Center of Condensed Matter Physics, Universidade de Brasilia,
Caixa Postal 04667, 70910-900 Brasilia, DF, Brazil}
\author{Evgueny Kochetov}
\affiliation{Bogoliubov Theoretical Laboratory, Joint Institute for Nuclear Research,
141980 Dubna, Russia}
\author{Marcin Mierzejewski}
\affiliation{Department of Theoretical Physics, Institute of Physics, University of Silesia, 40--007 Katowice, Poland}
\begin{abstract}
Within the $t$--$J$ model Hamiltonian
we present a RVB mean field theory directly in terms of dopant
particles. We apply this theory to $\mathrm{Na}_{x}\mathrm{CoO_{2}}\cdot y%
\mathrm{H_{2}0}$ and show that
the resulting
phase diagram $T_c$ versus doping is in qualitative agreement
with the experimental results.
\end{abstract}
\pacs x 74.20.Mn, 74.20.-z
\maketitle

\section{Introduction}
The remarkable discovery of superconductivity in $\mathrm{Na}_{x}\mathrm{%
CoO_{2}}\cdot y\mathrm{H_{2}0}$ for $x=0.35$ and $y=1.30$ by Takada \textit{%
et al.} \cite{takada} attracted a lot of attention. The
experimental findings indicate several striking similarities
between the
cobaltates and the cuprates. $\mathrm{Na}_{x}\mathrm{CoO_{2}}\cdot y\mathrm{%
H_{2}0}$ can be viewed as a $2D$ Mott insulator. The Co atoms form a
triangular lattice but $\mathrm{Co}^{4+}$ is in a $s=\frac{1}{2}$
low spin state. The
transition temperature ($T_{c}$) is seen to decrease for both underdoped and
overdoped materials \cite{tomek} although for the cobaltates the maximum $%
T_{c}$ is much lower ($T_{c}\approx $ $5K$) and the optimal doping is twice
as large as in the cuprates. Finally, as one varies temperature and electron
concentration, apart from superconductivity, there are observed unusual
electronic properties \cite{foo} and clear hints that strong electronic
correlation is at work in both cases.

All this turns attractive the application of the resonating valence bond
(RVB) ideas to this new compound. Baskaran~\cite{baskaran} was the first to
present qualitative arguments in favor of the RVB approach for the
cobaltates. Soon after that Kumar and Shastry \cite{shastry} as well as Lee
and coworkers \cite{lee1,lee2} presented their first estimates for the mean
field (MF) phase diagram, $T_{c}$ versus doping, in a RVB\ framework. In the
cobaltates the $\mathrm{CoO}_{2}$ layers are arranged in a triangular lattice which
naturally exhibits considerable magnetic frustration.
This not only brings modifications to the symmetry of the
resulting superconducting state, as pointed out
by others,~\cite{baskaran,shastry,lee1,lee2,baskaran1} but makes
the the application of a MF RVB, to reproduce the experimental findings
related to the superconductivity of $\mathrm{Na}_{x}\mathrm{%
CoO_{2}}\cdot y\mathrm{H_{2}0}$, even more challenging. The reasons
for that are as follows. The maximum $T_{c}$ for  the $\mathrm{CoO}_{2}$
layers occurs for doping values nearly twice as large as in the
cuprates. This might be indicative that the phase fluctuations of
the superconducting (SC) order parameter could be much too strong
for the stability of the MF RVB state. Moreover, in the standard
approach low doping is always favored and makes even harder a more
quantitative agreement with experiment in the case of the
cobaltates.

Within the original Baskaran-Zou-Anderson (BZA) MF
approximation~\cite{bza} a non-zero value of the RVB MF order
parameter (OP) does not by itself imply superconductivity. The true
SC OP in their approach is essentially taken as a product of a
spinon (a spin-1/2 neutral fermion) pairing OP  and a bose
condensation factor for the holons (spin-0 charged
bosons),~\cite{ubbens} following the more conventional slave-boson
approximation. The phase of the OP accounts for the fluctuations
which drives $T_c$ to zero at zero doping. The bose condensation
temperature for the holons is estimated separately and the region in
which both the spin-pairing OP and the mentioned holon bose factor
are non-zero determines the resulting RVB superconducting
phase.~\cite{shastry} All this seems to indicate that the standard
MF RVB basic ingredients -- the lattice spin singlet pairs -- which
might indeed be appropriate to describe the physics at very low
dopings is not the best starting point to address the
superconducting regime at higher doping values.

To overcome those limitations we present a RVB MF scheme which
takes direct account of the dopant particles themselves and treats
the non-double occupancy (NDO) constraint beyond the conventional
slave-boson mean field approximation. As will be demonstrated  by
this and later works this RVB representation is most suitable to
deal with cases in which strongly correlated electronic
superconductivity is manifest in both low and high doping regimes.

In this first work we apply our method to the $\mathrm{CoO}_{2}$
superconductors. Our starting point is the $t$-$J$ model on a triangular
lattice. In doing that we follow the arguments which consider the
$3d$ levels of the $\mathrm{Co}^{4+}$ ions being crystal field split in the
$\mathrm{CoO}_2$ layers producing singly occupied non-degenerate spin-$1/2$
$d_{z^2}$ orbitals. Those orbitals are directly associated with the
singlet states in our $t$-$J$ model representation based on the Hubbard
$X$ operators. The use of that representation will allow us to go
beyond the conventional treatment of the NDO
constraint. The $X$ operators are later given in a convenient
coherent-state path-integral representation. The resulting variables
in this representation are naturally split into bosonic and
fermionic degrees of freedom. The bosonic modes correspond to SU(2)
spin excitations while the fermion variables are spinless and
describe U(1) charge excitations instead. Combining those spinon
amplitudes and spinless fermion parameters together we then
construct appropriate fermionic fields which carry both spin and
charge degrees of freedom and can be directly related to the dopant
carriers in the $t$-$J$ model. We are able in this way to take a direct
account of the doping dependence of the critical superconducting
temperature preserving all the symmetry properties of the $t$-$J$
Hamiltonian.

We reformulate the RVB theory of the SC phase entirely in terms of those
quasiparticle states and use this scheme initially to describe the
superconducting properties of the cobaltates. We find qualitatively good
agreement with experiment and we are able   to reproduce
the observed dome structure of the $T_c$ versus doping phase diagram for those
materials in a RVB MF framework. The application of our RVB method to the
cuprates will be presented in a subsequent work.

\section{$t$-$J$ Hamiltonian and the NDO constraint}

We start by expressing the $t$-$J$ Hamiltonian \cite{spalek}
\begin{equation}
H_{t-J}=-t\sum_{ij\sigma }c_{i\sigma}^{\dagger}c_{j\sigma}+h.c.
+J\sum_{ij}\left( \overrightarrow{Q}_{i}\overrightarrow{Q}_{j}-\frac{1%
}{4}n_{i}n_{j}\right) ,  \label{1.1}
\end{equation}%
with
the NDO constraint, $\sum_{\sigma
}n_{i\sigma \text{ }}\leq 1,$ in terms of the Hubbard operators,~\cite{h}
$$X_{i}^{\sigma 0}=c_{i\sigma }^{+}\left( 1-n_{i-\sigma }\right),
\quad n_{i-\sigma }n_{i\sigma }=0.$$ Here $c_{i\sigma }$ is the
electron annihilation operator at site $i$ with the spin projection
$\sigma =\uparrow \downarrow $, $n_{i\sigma \text{ }}=c_{i\sigma
}^{+}c_{i\sigma },$ and the ${\vec Q}$'s are the corresponding
electron spin operators. In terms of these operators the local
NDO constraint holds rigorously and the
$t$-$J$ model becomes
\begin{equation}
H_{t-J}=-t\sum_{ij\sigma }X_{i}^{\sigma 0}X_{j}^{0\sigma
}+h.c.+J\sum_{ij}\left( \overrightarrow{Q}_{i}\overrightarrow{Q}_{j}-\frac{1%
}{4}n_{i}n_{j}\right) ,  \label{1.2}
\end{equation}%
where the electron spin operator now reads $\overrightarrow{Q_{i}}=\frac{1}{2}%
\sum_{\sigma \sigma ^{\prime }}X_{i}^{\sigma 0}\overrightarrow{\tau }%
_{\sigma \sigma ^{\prime }}X_{i}^{0\sigma ^{\prime }}$, with the $\overrightarrow{%
\tau }^{\prime}$s being Pauli matrices.

Fermionic operators $X_{i}^{\sigma 0}$ project the electron creation
operators onto a space spanned by the basis $\left\{ |0\rangle
_{i},|\sigma \rangle _{i}\right\} $ and take the form $X_{i}^{\sigma
0}=|\sigma \rangle _{i}\langle 0|_{i}.$ Together with the bosonic
generators, $X^{\sigma\sigma'}_i=|\sigma\rangle_i \langle\sigma'|_i$
the full set of operators $X_i^{ab},\, a,b=0,\uparrow,\downarrow$
forms, on every lattice site, a basis of the fundamental
representation of the semisimple doubly graded Lie algebra su$(2|1)$
given by the (anti)commutation relations
$$\{X^{ab}_i,X^{cd}_j\}_{\pm}=(X_i^{ad}\delta^{bc}\pm X^{bc}_j
\delta^{ad})\delta^{ij},$$ where the $(+)$ sign should be used only
when both operators are fermionic.

Since su$(2|1)$ can be viewed as a supergeneralization of the
conventional spin su(2) algebra, the $t$-$J$ Hamiltonian appears as a
superextension of the Heisenberg magnetic Hamiltonian, with a hole
being a superpartner of a su(2) magnetic excitation.~\cite{w} This
superalgebra can also be thought of as a natural generalization of
the standard fermionic algebra spanned by generators $c_{\sigma
}^{+},$$c_{\sigma },$ and unity $I,$ to the case where the fermionic
operators are subject to the NDO constraint. The
incorporation of this constraint manifests itself in more
complicated commutation relations between $X$ operators in
comparison with those produced by the conventional fermionic
operators. Note that the Gutzwiller projection $P_G=\prod_{i}\left(
1-n_{i\sigma }n_{i-\sigma }\right)$ that excludes the doubly
occupied state $|\uparrow \downarrow \rangle $ is equivalent to the
Hubbard operator representation, since $P_Gc_{i\sigma }^{+}c_{j\sigma}P_G
=X_{i}^{\sigma 0}X_{j}^{0\sigma }$ .

Note also that the occupation constraint is different for the hole
and electron doping. To treat them in a unique way we perform, for
electron dopings, a canonical particle-hole transformation
$c_{i\sigma }\rightarrow c_{i-\sigma }^{+}$ that restores the
non-double occupancy constraint but reverses the sign of $t$. Using
then the Hubbard operator representation in terms of the transformed
$c$-operators we again arrive at Eq.(\ref{1.2}) with, however, $t\to
-t$. Although the $\mathrm{CoO_{2}}$ layer is an electron doped Mott
insulator we shall for convenience formally deal with the more
familiar case of hole doping making the necessary changes only at
the end of our work.

Since the $X$ operators are generators of the su$(2|1)$ superalgebra
we are lead naturally to employ the su$(2|1)$ coherent-state
path-integral representation of the $t$-$J$ partition function. There
are a few rationales to do that. First, this provides a mathematical
setting well adjusted to address the $t$-$J$ model with the crucial
NDO constraint naturally built in the formalism
from the very beginning. Second, within the su$(2|1)$ path-integral
representation the associated effective $t$-$J$ action lives on a
natural classical phase space of the $t$-$J$ model -- the SU$(2|1)$
homogeneous compact manifold, CP$^{1|1}$ (see below). The group
SU$(2|1)$ acts on the CP$^{1|1}$ manifold as a group of canonical
transformations in a way that the transformation properties of the
basic fields -- the local coordinates on CP$^{1|1}$ -- can be easily
found. Third, these coordinates are naturally split into bosonic and
fermionic degrees of freedom. In the context of the $t$-$J$ model the
bosonic fields correspond to the SU(2) spin excitations whereas the
fermionic ones are spinless and may be used to describe the U(1)
charge excitations. This provides a natural setting to implement the
spin-charge separation inherent in the spin liquid phase at least in
1D. Finally, the transformation properties of the CP$^{1|1}$
coordinates under global SU(2)$\times$ U(1) rotations -- the exact
symmetry of the $t$-$J$ Hamiltonian --  imply that their certain
combinations transform in the linear spinor representations of SU(2)
and may therefore be used to describe fermionic quasiparticle
excitations that carry both the charge and spin quantum numbers. We
show that such quasiparticles arise as the dopant particles in the
$t$-$J$ model relevant for describing the SC phase. In particular, we
formulate the RVB theory of the SC phase of the $t$-$J$ model directly
in terms of the dopant particles and apply it to describe the SC in
the cobaltates.

Some earlier attempts have also been made to apply supersymmetry
to study many--fermion interacting Hamiltonians as well as the $t$--$J$ and
related models. A linearization scheme for the general Hamiltonian of an
interacting fermion system has been proposed in Ref. ~\onlinecite{solomon}.
A hierarchy of spectrum--generating algebras and superalgebras
including su$(2|1)$ results from such a new mean--field treatment.
A supersymmetric representation of the
Hubbard operator which unifies the slave--boson and
slave--fermion representation
into a single U$(1|1)$ gauge theory has been developed in Ref.
~\onlinecite{pepin}.
Such a representation makes unnecessary the choice between a bosonic
and fermionic
spin and is most suitable to describe the coexistence of strong magnetic
correlations within a paramagnetic phase. Besides, it has been demonstrated
\cite{coleman} that thus defined supersymmeric Hubbard operators prove to
be very efficient in treating the physics of the infinite U Hubbard model.

\section{su$(2|1)$ coherent states and path integral}

The normalized su$(2|1)$
coherent state (CS) associated with the 3D fundamental representation
takes the form
\begin{equation}
|z,\xi\rangle=(1+\bar{z}z
+\bar{\xi}\xi)^{-1/2}\exp\left(zX^{\downarrow\uparrow}+\xi
X^{0\uparrow}\right)|\uparrow\rangle,\label{eq:cs}
\end{equation} where $z$ is a complex number, and $\xi $ is a complex
Grassmann parameter. The set $(z,\xi)$ can be thought of as local
coordinates of a given point on CP$^{1|1}$. This supermanifold
appears as a $N=1$ superextension of a complex projective plane, or
ordinary two--sphere, CP$^1=S^2$, to accommodate one extra complex
Grassmann parameter.~\cite{k2} At $\xi =0$, the su$(2|1)$ CS reduces
to the ordinary su(2) CS, $|z,\xi=0\rangle\equiv |z\rangle$
parametrized by a complex coordinate $z \in CP^1$. Note that the
classical phase space of the Hubbard operators, CP$^{1|1}$, appears
as a $N=1$ superextension of the CS manifold for the su(2) spins.

As is well known the key point in constructing the coherent--state path--integral
representation of a partition function
is the resolution of unity or, equivalently, the completeness relation for
the coherent states.
In terms of the normalized set of states~(\ref{eq:cs}) it takes the form
$$\int d\mu_{SU(2|1)}|z\xi\rangle\langle z\xi|=I,$$
where
$$d\mu_{SU(2|1)}=\frac{d\bar zdz}{2\pi i}\,\frac{d\bar\xi d\xi}{1+|z|^2+\bar\xi\xi}$$
stands for the SU$(2|1)$ invariant measure on the coherent-state manifold,
CP$^{1|1}$=SU$(2|1)$/U$(1|1),$ and $I$ is the identity operator in the projected
Hilbert space.
Explicitly, we have
\begin{eqnarray} \int d\mu|z\xi\rangle\langle z\xi| =
\int\frac{d\bar zdzd\bar\xi d\xi}{2\pi i(1+|z|^2+\bar\xi\xi)}|z\xi\rangle
\langle z\xi| && \nonumber  \\
=\int\frac{d\bar zdzd\bar\xi d\xi}{2\pi i(1+|z|^2+\bar\xi\xi)}\,
\frac{1}{(1+|z|^2+\bar\xi\xi)} && \nonumber \\
\times \left(|\uparrow\rangle\langle \uparrow|+
|z|^2|\downarrow\rangle\langle\downarrow|+\xi\bar\xi|0\rangle\langle 0|\right) && \nonumber \\
=|\uparrow\rangle\langle \uparrow|+
|\downarrow\rangle\langle\downarrow|+|0\rangle\langle 0|\equiv I . && \nonumber
\end{eqnarray}

In the basis $|z,\xi\rangle=\prod_j |z_j,\xi_j\rangle$,
the $t$--$J$ partition function takes the form of
the su$(2|1)$ CS phase-space path integral,
\begin{equation}
Z_{t-J}=tr\ \exp (-\beta H_{t-J})=\int_{CP^{1|1}} D\mu _{SU(2|1)}(z,\xi )\
e^{S_{t-J}}, \label{2.1}
\end{equation}%
where
$$D\mu _{SU(2|1)}(z,\xi )=\prod_{j,t}\frac{d\bar z_j(t)dz_j(t)}{2\pi i}\,
\frac{d\bar\xi_j(t)d\xi_j(t)}{1+|z_j|^2+\bar\xi_j\xi_j}$$ stands for
the SU$(2|1)$ invariant measure with the boundary conditions,
$z_j(0)=z_j(\beta),\, \xi_j(0)= -\xi_j(\beta).$ The $t$-$J$ effective
action on CP$^{1|1}$ now reads $S_{t-J}=-\int_0^{\beta}\langle
z,\xi|d/dt+ H_{t-J}|z,\xi\rangle dt$, which gives
\begin{eqnarray}
S_{t-J}&=&\frac{1}{2}\sum_j\int_0^{\beta}\frac{\dot{\bar z_j}z_j-\bar
z_j\dot z_j
+\dot{\bar\xi_j}\xi_j-\bar\xi_j\dot\xi_j}{1+|z_j|^2+\bar\xi_j\xi_j}dt \nonumber \\
&&-
\int_0^{\beta}H^{cl}_{t-J}dt. \label{2.2}\end{eqnarray}
The first
part of the action (\ref{2.2}) is a purely kinematical term that
reflects the geometry of the underlying phase space while the
classical image of the Hamiltonian (\ref{1.2}) becomes an average
value of $H_{t-J}$ over the su$(2|1)$ coherent states,
\begin{eqnarray}
H^{cl}_{t-J}&=&\langle z,\xi|H_{t-J}|z,\xi\rangle \nonumber \\
&=&-t\sum_{ij}\frac{\xi_i\bar\xi_j(1+z_j
\bar z_i) + h.c.}
{(1+|z_i|^2+\bar\xi_i\xi_i)(1+|z_j|^2+\bar\xi_j\xi_j)} \nonumber \\
&&+J\sum_{ij}\frac{-|z_i|^2-|z_j|^2+z_iz_j+
\bar z_i\bar z_j}
{(1+|z_i|^2+\bar\xi_i\xi_i)(1+|z_j|^2+\bar\xi_j\xi_j)}. \nonumber \\
\label{2.3}\end{eqnarray}

The fact that the electron system with the NDO constraint
lives on the compact manifold, supersphere
CP$^{1|1}$ can be explained as follows. Let us for a moment suppose
that the so-called slave-fermion representation for the electron
operators is used, i.e.,
\begin{equation}
c_{i\sigma }=f_ia_{i\sigma }^{\dagger},
\label{c}\end{equation}
where $f_i$ is
a on-site spinless fermionic operator, whereas $a_{i\sigma }$ is
the spinful boson. The NDO constraint now reads $%
\sum_{\sigma }a_{i \sigma }^{+}a_{i \sigma }+f^{+}_i f_i=1.$ Within the slave-fermion
path integral representation
\begin{equation}
Z_{t-J}=\int D\mu _{flat}\ e^{S_{t-J}(\overline{a}_{\sigma
},a_{\sigma },f)}, \label{2.4}
\end{equation}
with the integration measure $D\mu
_{flat}=\prod_{i}D\overline{a}_{i\uparrow
}Da_{i\uparrow }D\overline{a}_{i\downarrow }Da_{i\downarrow }D\overline{f_{i}%
}Df_{i}$ , this constraint transforms into
\begin{equation}
\sum_{\sigma }\overline{a}_{i\sigma }a_{i\sigma }+\overline{f}_{i}f_{i}=1,
\label{2.5}
\end{equation}
with $a_{i\sigma }$ and $f_i$ standing now for complex numbers and
complex Grassmann parameters, respectively. Equation (\ref{2.5}) is
exactly that for the supersphere CP$^{1|1}$ embedded into a flat
superspase. Any mean-field treatment of (\ref{2.4}) should respect
this constraint, which, however, poses a severe technical problem.
If one however resolves this equation explicitly by making the
identifications
\begin{eqnarray}
a_{i\uparrow }&=&\frac{e^{i\phi_i}}{\sqrt{1+\overline{z}_iz_i
+\overline{\xi}_i\xi_i}},\
a_{i\downarrow }=\frac{z_ie^{i\phi_i}}{\sqrt{1+\overline{z}_iz_i
+\overline{\xi}_i\xi_i}},\
\nonumber \\
f_i&=&%
\frac{\xi_i e^{i\phi_i}}{\sqrt{1+\overline{z}_iz_i+\overline{\xi}_i\xi}_i},
\label{2.6}
\end{eqnarray}
one can further treat the variables $z_i,\xi_i $ as if they were indeed
free of any constraints.

Note that the electron operator~(\ref{c}) is invariant under a local gauge
transformation,
$$a_{i\sigma}\to a_{i\sigma}e^{i\theta_i},\quad f_i\to f_ie^{i\theta_i},$$ or
equivalently, under the change $\phi_i\to \phi_i+\theta_i$. This gauge symmetry
is a consequence of the redundancy of parameterizing the electron operator in terms
of the auxiliary boson/fermion fields. In contrast, the $su(2|1)$ projected
coordinates
$$z_i=a_{i\downarrow}/a_{i\uparrow},\quad \xi_i=f_i/a_{i\uparrow}$$ are seen
to be manifestly gauge invariant.
The domain of the flat measure in (\ref{2.4}) that involves
the spin up bosonic fields can be rewritten
at every lattice site
as $D\bar a_{i\uparrow}Da_{i\uparrow}=D|a_{i\uparrow}|^2D\phi_i$.
The $|a_{i\uparrow}|^2$ field can easily be integrated out from
eq.(\ref{2.4}) because of
the constraint (\ref{2.5}). Since the t-J action is U(1) gauge invariant and hence
independent of $\phi_i$,  the integration
over $\phi_i$ results in merely some numerical factor that can be taken care of
by a proper normalization of the partition function.
For the remaining integration we have (the site dependence for the moment being
suppressed),
$$Da_{\downarrow}D\bar a_{\downarrow}DfD\bar f= {\rm sdet}
\|\frac{\partial(a_{\downarrow},\bar a_{\downarrow},f,\bar f)}
{\partial(z, \bar z, \xi,\bar\xi)}\|\,DzD\bar zd\xi D\bar\xi.$$
The Jacobian of the change of the supercoordinates appears as
a superdeterminant of the transformation matrix~\cite{berezin}
\begin{eqnarray}
{\rm sdet}\|\frac{\partial(a_{\downarrow},\bar a_{\downarrow},f,\bar f)}
{\partial(z, \bar z, \xi,\bar\xi)}\| &= &
{\rm sdet}\left(
\begin{array}{cc}A&B\\C&D
\end{array}\right) \nonumber \\
&:= &{\rm det} (A-BD^{-1}C) {\rm det} D^{-1}. \nonumber
\end{eqnarray}
Here
$$A=\left(
\begin{array}{cc}\frac{\partial a_{\downarrow}}{\partial z}&
\frac{\partial a_{\downarrow}}{\partial \bar z}\\
\frac{\partial \bar a_{\downarrow}}{\partial  z} &
\frac{\partial \bar a_{\downarrow}}{\partial \bar z}\end{array}\right),\quad
B=
\left(
\begin{array}{cc}\frac{\partial a_{\downarrow}}{\partial \xi}&
\frac{\partial a_{\downarrow}}{\partial \bar\xi}\\
\frac{\partial \bar a_{\downarrow}}{\partial  \xi} &
\frac{\partial \bar a_{\downarrow}}{\partial \bar\xi}\end{array}\right),$$
$$C=\left(
\begin{array}{cc}\frac{\partial f}{\partial z}&
\frac{\partial f}{\partial \bar z}\\
\frac{\partial \bar f}{\partial  z} &
\frac{\partial \bar f}{\partial \bar z}\end{array}\right),\quad
D=
\left(
\begin{array}{cc}\frac{\partial f}{\partial \xi}&
\frac{\partial f}{\partial \bar\xi}\\
\frac{\partial \bar f}{\partial  \xi} &
\frac{\partial \bar f}{\partial \bar\xi}\end{array}\right),$$
with the derivatives with respect to the Grassmann parameters
$\xi$ and $\bar\xi$ being understood to be the right ones.

Evaluating the superdeterminant
$${\rm sdet} \|\frac{\partial(a_{\downarrow},\bar a_{\downarrow},f,\bar f)}
{\partial(z,\bar z, \xi,\bar\xi)}\|=\frac{1}{1+|z|^2+{\bar\xi}\xi}$$ and
substituting of (\ref{2.6}) into (\ref{2.4}) we are led to the su(2%
\mbox{$\vert$}%
1) path-integral representation of $Z_{t-J}$ given by (\ref{2.1}).
Note that the U(1) gauge field $\phi_i$ drops out from representation
~(\ref{2.1}). An attempt at decoupling the physical electron as
a U(1) gauge invariant "dressed" holon and spinon has been made
in Ref. \onlinecite{gauge}.

Geometrically, the set $(z,\xi)$ appears as local (inhomogeneous)
coordinates of a point on the supersphere defined by equation
(\ref{2.5}).
Representation (\ref{2.1})-(\ref{2.3}) rigorously
incorporates the local NDO constraint at the
apparent expense of a more
complicated compact phase space for the projected electron operators.

\section{Symmetry}

At the supersymmetric point, $J=2t,$ the $t$--$J$ model Hamiltonian
is known to exhibit a global SU$(2|1)$ symmetry. Away from that
point this symmetry reduces to SU(2)$\times$ U(1)$\subset$
SU$(2|1)$. This symmetry group acts on a point $(z(t),\xi(t))\in
CP^{1|1}$ in a way that,
\begin{eqnarray}
z(t)\rightarrow z_{g}(t)=\frac{uz(t)+v}{-\overline{v}z(t)+\overline{u}}, && g\in
\mathrm{SU(2)\times U(1)}, \nonumber \\
\xi (t)\rightarrow \xi_{g}(t)
=\frac{e^{i\theta}\xi (t)}{-\overline{v}z(t)+\overline{u}},
&&  \label{3.1}
\end{eqnarray}
where the group parameters are to be taken to be site independent:
\begin{eqnarray}
\left(\begin{array}{ll}
u & v \\
-\overline{v} & \overline{u}%
\end{array}
\right) \in \mathrm{SU(2)},\quad e^{i\theta}\in \mathrm{U(1)}.
\end{eqnarray}

It can easily be checked that both the SU$(2|1)$ measure and the
effective action (\ref{2.2}) are invariant under the group
transformations (\ref{3.1}), so that the representation of the
partition function (\ref{2.1}) remains intact. Notice that
(\ref{3.1}) appears as a covariant reparametrization of CP$^{1|1}$.
However, one can in principle
employ any other reparametrization, not necessarily of the form of the
SU(2\mbox{$\vert$}1)
action on CP$^{1|1}$. We are interested in the one that decouples the
SU(2\mbox{$\vert$}1) measure factor into the SU(2) spin and the U(1) spinless fermion
measures,
\begin{eqnarray}
D\mu _{SU(2)}(\bar z,z)&=&\prod_{j,t}\frac{d\bar z_j(t)dz_j(t)}
{2\pi i(1+|z_j(t)|^2)^2},\nonumber \\
D\mu _{U(1)}(\bar\xi,\xi)&=&\prod_{j,t}
d\bar\xi_j(t)d\xi_j(t), \nonumber
\end{eqnarray}
 respectively.

Such a reparametrization can be taken to be
\begin{equation}
z\rightarrow z,\ \xi \rightarrow \xi \sqrt{1+|z|^{2}}.  \label{3.2}
\end{equation}
Up to an inessential factor which redefines a chemical potential, we get
\begin{equation}
D\mu _{su(2|1)}\rightarrow D\mu _{su(2)}(\overline{z},z)\times D\mu _{u(1)}(%
\overline{\xi },\xi ),  \label{3.3}
\end{equation}
and the effective action becomes
\begin{eqnarray}
S_{t-J}\rightarrow S_{t-J}=\frac{1}{2}\sum_{i}\int_{0}^{\beta }\frac{%
\overline{\dot{z}}_{i}z_{i}-\overline{z}_{i}\dot{z}_{i}}{1+\overline{z}_{i}z_{i}}%
(1-\overline{\xi _{i}}\xi _{i})dt && \nonumber \\
+\frac{1}{2}\sum_{i}\int_{0}^{\beta }%
\overline{(\dot{\xi}}_{i}\xi _{i}-\overline{\xi }_{i}\dot{\xi}%
_{i})dt-\int_{0}^{\beta }\widetilde{H}_{t-J}^{cl}(t)dt, &&  \label{3.4}
\end{eqnarray}
with
\begin{eqnarray}
&&\widetilde{H}_{t-J}^{cl}=
-t\sum_{ij}(\xi _{i}\overline{\xi }_{j}\langle
z_{i}|z_{j}\rangle +h.c.)  \nonumber  \\
&&+\frac{J}{2}\sum_{ij}\left( |\langle
z_{i}|z_{j}\rangle |^{2}-1\right) \left( 1-\overline{\xi }_{i}\xi
_{i}\right) \left( 1-\overline{\xi }_{j}\xi _{j}\right).   \label{3.5}
\end{eqnarray}
Here $\langle z_{i}|z_{j}\rangle$ stands for an inner product of the su(2)
coherent states, $$\langle z_{i}|z_{j}\rangle =\frac{1+\overline{z}%
_{i}z_{j}}{\sqrt{(1+|z_{j}|^{2})(1+|z_{i}|^{2})}}.$$

From eqs.~(\ref{3.1}) one can infer the transformation
properties of the new CP$^{1|1}$ coordinates (\ref{3.2}) under a global SU(2)$%
\times $U(1) action:
\begin{eqnarray}
z(t)\rightarrow
z_{g}(t)&=&\frac{uz(t)+v}{-\overline{v}z(t)+\overline{u}},
\nonumber \\
\xi(t)\rightarrow \xi _{g}(t)&=&e^{i\phi _{g}+i\theta }\xi (t),\   \label{3.6}
\end{eqnarray}
where $$i\phi _{g}=\ln \sqrt{\frac{-v\overline{z}+u}{-\overline{v}z+\overline{%
u}}}.$$ Note also that$\ |z\rangle \rightarrow |z\rangle
_{g}=e^{-i\phi _{g}}|z_{g}\rangle.$ It can be straightforwardly
checked that both the measure and the t-J action (\ref{3.4}) remain
invariant, under such an action of SU(2)$\times $U(1).

The following remarks are needed at this stage. First, in spite of
the fact that the function $i\phi _{g}$ bears a site-dependence
through the $z_i$ fields, the transformation (\ref{3.6}) is a global
one: the group parameters $(u,v)$ are site-independent. Second,
although the measure factor gets decomposed into the su(2) spin and
spinless fermion pieces, the underlying phase space is not reduced
into a direct product of the classical spin and a flat fermionic
phase spaces. The function $\phi _{g}$ that enters the
transformation law for the fermions also depends on the spinon
coordinates, $z_i(t).$ Besides, the symplectic one-form (kinetic
term) in the effective action (\ref{3.4}) is not a simple sum of
purely fermionic and spin contributions. This means physically that,
in general, the corresponding field excitations are not independent
of each other. In the other words, the spin-charge separation does
not merely reduce to a simple $(z,\,\xi)$ representation, and
should, in fact, be described by nonlocal `string` excitations to be
constructed out of the basic $(z,\,\xi)$ fields.

\section{Effective action}

The spinon amplitudes $z_{i}(t)$ and the spinless fermion
parameters, $\xi _{i}(t),$ are in fact related to each other by the
SU(2) transformation laws (\ref{3.6}). From this it follows that
we can construct classical images\cite{image} for the operators that
describe doped holes. In this respect we make the following ansatz:
\begin{eqnarray}
&&\Psi _{\downarrow }=\frac{-\xi }{\sqrt{1+|z|^{2}}},\ \overline{\Psi }%
_{\downarrow }=\frac{-\overline{\xi }}{\sqrt{1+|z|^{2}}},  \nonumber \\
&&\Psi _{\uparrow }=\frac{\overline{z}\xi }{\sqrt{1+|z|^{2}}},\ \overline{%
\Psi }_{\uparrow }=\frac{z\overline{\xi }}{\sqrt{1+|z|^{2}}}.  \label{4.0}
\end{eqnarray}%
It then follows that $\ \overline{\Psi }_{\uparrow }\Psi _{\uparrow }+%
\overline{\Psi }_{\downarrow }\Psi _{\downarrow }=\overline{\xi }\xi =%
\widehat{\delta }^{cl},$ where $\widehat{\delta }^{cl}$ stands for a
classical image of the hole--number operator $\widehat{\delta }=1-\widehat{n}%
_{e}=1-\sum_{\sigma }X^{\sigma \sigma }.$ Therefore the resulting
fermionic amplitudes describe the propagation of doped holes
restricted to the NDO constraint. In view of the group
transformations SU(2)$\times $U(1) for $z(t)$ and $\xi(t)$ the $\Psi
_{\sigma }$ amplitudes transform in a \emph{linear} spinor
representation of SU(2) as true fermionic amplitudes. Namely,
\begin{eqnarray}
\left(\begin{array}{l}\Psi_{\uparrow}\\ \Psi_{\downarrow}
\end{array}\right)
\rightarrow
\left(\begin{array}{ll}
\overline{u} & -\overline{v} \\
v & u
\end{array}\right)
\left(\begin{array}{l}\Psi_{\uparrow}\\ \Psi_{\downarrow}
\end{array}\right)
\label{4.1}
\end{eqnarray}

In terms
of the $\Psi _{\sigma }$ and $z$ amplitudes we get the corresponding \textit{%
exact} representation of the $t$--$J$ partition function,
\begin{eqnarray}
Z_{t-J} &=&\int D\mu _{SU(2)}(\overline{z},z)D\mu _{U(1)}(\overline{\Psi }%
,\Psi )\exp S  \nonumber \\
&\times& \prod_{i}\delta \left( \frac{\Psi _{\uparrow i}+\overline{z}%
_{i}\Psi _{\downarrow i}}{\sqrt{1+|z_{i}|^{2}}}\right) \delta \left( \frac{%
\overline{\Psi }_{\uparrow i}+z_{i}\overline{\Psi }_{\downarrow i}}{\sqrt{%
1+|z_{i}|^{2}}}\right),  \label{4.2}
\end{eqnarray}%
where
\begin{equation}
S=S_{kin}-\int_{0}^{\beta }{\tilde H}_{t-J}^{cl}(t)dt.
\label{4.3}\end{equation} The SU(2) invariant product of the
$\delta$ - functions ensures the preservation of the correct number
of degrees of freedom. The square roots in the $\delta $- function
arguments come from the evaluation of the Jacobian. In this new
dopant carrier representation the kinetic term,
\begin{eqnarray}
S_{kin}&=&\frac{1}{2}\sum_{\sigma i}\int_{0}^{\beta }\left( \overline{\dot{%
\Psi}}_{\sigma i }\Psi _{\sigma i }-\overline{\Psi }_{\sigma i
}\dot{\Psi}_{\sigma i }\right) dt \nonumber \\
&&+\frac{1}{2}\sum_{i}\int_{0}^{\beta }\frac{\overline{\dot{z}}_{i}z_{i}-%
\overline{z}_{i}\dot{z}_{i}}{1+\overline{z}_{i}z_{i}}dt,
\label{4.4}
\end{eqnarray}
is nicely decoupled into purely fermionic
and spinon parts. It is clear that the fermionic symplectic
one--form (the first term in (\ref{4.4})) determines a standard
fermionic symplectic structure
$\sum_{\sigma }d\overline{\Psi }_{\sigma }\wedge d\Psi _{\sigma }$
which in turn determines the standard Poisson brackets relations  $\left\{ \overline{%
\Psi }_{\sigma },\Psi _{\sigma ^{\prime }}\right\} _{PB}=\delta
_{\sigma ,\sigma ^{\prime }},\left\{ \Psi _{\sigma },\Psi _{\sigma
^{\prime }}\right\} _{PB}=0.$ As a result the corresponding
operators $\Psi _{\sigma }^{+}\ ,\Psi _{\sigma ^{\prime }}$ describe
indeed well-defined fermionic excitations - in our case, doped
holes.

As a result, using the new fermion fields, the Hamiltonian that
corresponds to ${\tilde H}_{t-J}^{cl}$ takes the form
\begin{eqnarray}
H_{t-J}& =&t\sum_{ij\sigma }\Psi _{i\sigma }^{+}\ \Psi _{j\sigma
}+h.c.+J\sum_{ij}\left[ (\overrightarrow{S}_{i}\overrightarrow{S}_{j}-\frac{1%
}{4})\right. \nonumber \\
&
+&\left. (\overrightarrow{S}_{i}\overrightarrow{M}_{j}+\overrightarrow{S}%
_{j}\overrightarrow{M}_{i})+(\overrightarrow{M}_{i}\overrightarrow{M}_{j}-%
\frac{1}{4}\widehat{\delta }_{i}\widehat{\delta }_{j})\right],  \label{4.5}
\end{eqnarray}%
where we have dropped the tilda sign.
The components of the operator of spinon magnetic moment $\overrightarrow{S}$
are the su(2) generators in the $s=%
{\frac12}%
$ representation . Their classical images are the components of
$\vec{S}^{cl}=\langle z|\vec{S}|z\rangle$ with
$(\vec S^{2})_{cl}=\frac{3}{4}.$ The hole spin operator $\overrightarrow{M}=\frac{1}{2}\sum_{\sigma }\Psi _{\sigma }^{+}%
\overrightarrow{\tau }_{\sigma \sigma ^{\prime }}\Psi _{\sigma },\
\overrightarrow{M}^{2}=\frac{3}{4}\widehat{\delta }\left( 2-\widehat{\delta }%
\right) ,  \widehat{\delta }=\Psi _{\uparrow }^{+}\Psi _{\uparrow
}+\Psi _{\downarrow }^{+}\Psi _{\downarrow },$ transforms under
(\ref{3.6}), as a SU(2) vector while the total hamiltonian
(\ref{4.5}) is a SU(2) scalar. It can also be checked that the
electron spin moment is a linear combination of the above two operators: $%
\overrightarrow{Q}=\overrightarrow{S}+\overrightarrow{M}$. If we
integrate out the fields ${\Psi}_{\uparrow i},\overline{\Psi
}_{\uparrow i}$ in Eq.~(\ref{4.2}), with the help of the $\delta -$
functions, we will return to our initial representation as given in
Eq.~(\ref{3.4}).

Different parts of the Hamiltonian~(\ref{4.5}) can be associated
with the different phases of the t-J model. For a
half-filled band, Eq.~(\ref{4.5}) reduces to the Heisenberg antiferromagnet (AF), $H^{AF}_{t-J}=J\sum_{ij}(%
\overrightarrow{S}_{i}\overrightarrow{S}_{j}-\frac{1}{4}).$ Away
from half-filling in the lightly doped regime, where $\delta$ is
small enough, so that one can ignore a direct hole-hole interaction,
the Hamiltonian
\begin{eqnarray}
H^{AF-PG}_{t-J}& =&t\sum_{ij\sigma }\Psi _{i\sigma }^{+}\ \Psi
_{j\sigma
}+h.c.+J\sum_{ij}\left[ (\overrightarrow{S}_{i}\overrightarrow{S}_{j}-\frac{1%
}{4})\right. \nonumber \\
&+&\left. (\overrightarrow{S}_{i}\overrightarrow{M}_{j}+\overrightarrow{S}%
_{j}\overrightarrow{M}_{i})\right], \label{4.6}\end{eqnarray}%
is able to describe the AF - pseudogap (PG) transition on a square
lattice. Accordingly, at higher doping, the Hamiltonian
\begin{eqnarray}
H^{PG-SC}_{t-J}& =&t\sum_{ij\sigma }\Psi _{i\sigma }^{+}\ \Psi
_{j\sigma }+h.c.
+J\sum_{ij}\left[\overrightarrow{S}_{i}\overrightarrow{M}_{j}\right. \nonumber \\&+&\left.\overrightarrow{S}%
_{j}\overrightarrow{M}_{i}+(\overrightarrow{M}_{i}\overrightarrow{M}_{j}-%
\frac{1}{4}\widehat{\delta }_{i}\widehat{\delta }_{j})\right],
\label{4.7}
\end{eqnarray}%
is appropriate for the pseudogap - superconductor boundary region of
the t-J phase diagram.

The PG phase itself can be described within our approach by a simple
SU(2) invariant spinon-fermion interaction,
\begin{eqnarray}
H^{PG}_{t-J}& =&t\sum_{ij\sigma }\Psi _{i\sigma }^{+}\ \Psi
_{j\sigma
}+h.c.\nonumber \\
&+& J\sum_{ij}\left[\overrightarrow{S}_{i}\overrightarrow{M}_{j}+\overrightarrow{S}%
_{j}\overrightarrow{M}_{i}\right]. \label{4.8}\end{eqnarray}%
To see this, one can recast the Hamiltonian~(\ref{4.8}) into the
form of the phenomenological boson - fermion model (BFM) which is
known to capture the main observable characteristic of the PG
phenomenon, namely the reduction of the fermionic density of states
at the Fermi level~\cite{rob}. Using the Holstein-Primakoff (HP)
representation of the spin operators on the bipartite lattice
$L=A\bigoplus B$,

$$S^z_{i}=1/2 -b^{\dagger}_ib_i,\quad S^{+}_i=-b_i,\quad
S^{-}_i=-b^{\dagger}_i,\quad i\in A,$$
$$S^z_{i}=1/2 -b^{\dagger}_ib_i,\quad
S^{+}_i=b_i,\quad S^{-}_i=b^{\dagger}_i,\quad i\in B,$$ where
$[b,b^{\dagger}]=1$, and performing the following unitary
transformation of the fermionic operators

$$\Psi_{\uparrow i}\to \Psi^{\dagger}_{\uparrow i}, \, i\in A, \quad
\Psi_{\uparrow i}\to -\Psi^{\dagger}_{\uparrow i}, \, i\in B,$$ one
is led to the BFM - type Hamiltonian

\begin{eqnarray}
H^{PG}_{t-J}&\to & H^{BFM}=t\sum_{ij}\Psi _{i\sigma }^{+}\ \Psi
_{j\sigma }+h.c.- J\sum_{i} \Psi^{\dagger}_{i\sigma}
\Psi_{i\sigma}\nonumber\\&-& 2J\sum_{i}b_i^{\dagger}b_i +
J\sum_{ij}(\Psi_{\uparrow
j}\Psi_{\downarrow j}b^{\dagger}_i +h.c)\nonumber \\
&+& \lambda\sum_{i}(2b_i^{\dagger}b_i+\Psi^{\dagger}_{i\sigma}
\Psi_{i\sigma}-2),
 \label{4.9}\end{eqnarray}%
with the  implied summation over $\sigma$. The Lagrange multiplier
$\lambda$ has been introduced to enforce the constraint which
assures the vanishing of the total electron spin projection,
$<2Q^z>=0$. Notice that due to the global U(1) invariance the
conditions  $<Q^{\pm}>=0$ are automatically satisfied. While both
the conventional BFM and $H_{t-J}^{PG}$. Hamiltonians possess the
same global symmetry, namely SU(2)$\times$ U(1), with the $SU(2)$
group describing of the rotation of the spinor fields, the origin of
the U(1) symmetry is different in the two cases. In the standard
BFM, the global U(1) symmetry corresponds to the conservation of the
total charge of bosons and fermions. Here the U(1) symmetry group
appears just as a subgroup of the explicitly broken (by the HP
representation) total spin rotation group, generated by the
operators $\vec Q´s$. Therefore, despite the formal similarity
between those two effective models, the physical contents of the
standard BFM and the representation (\ref{4.9}) are indeed different from each other .

Note, finally, that our discussed hierarchy of effective t-J
Hamiltonians is basically qualitative in the sense that the
constraint imposed by the $\delta$-functions might change their
detailed forms. While this constraint does not seem to be of the
crucial importance at very low density of dopant carriers, it
definitely becomes more important as $\delta$ increases, and this
may in turn substantially affect the final form of the effective
interactions. However, since the global SU(2) invariance of the t-J
Hamiltonian~(\ref{4.5}) is not affected by the constraint, it is
plausible to suggest that those changes will at most, at the
mean-field level, result merely in a renormaization of the
Hamiltonian parameters similar to what happens to the t and J
parameters in the mean-field Plain Vanilla theory~\cite{pvt}. We
intend to address these problems in more detail in a separate
publication.

\section{SC phase}

\begin{figure}[tbp]
\vspace*{3mm} \includegraphics[width=8.5cm]{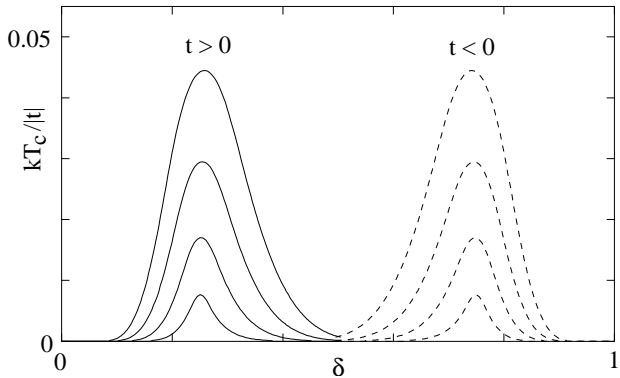}
\caption{$T_c(\protect\delta)$ for negative and positive $t$. The curves
from the bottom to the top correspond to $J/|t|=$0.6, 0.8, 1, 1.2.}
\label{result}
\end{figure}

\begin{figure}[tbp]
\vspace*{3mm} \includegraphics[width=8.5cm]{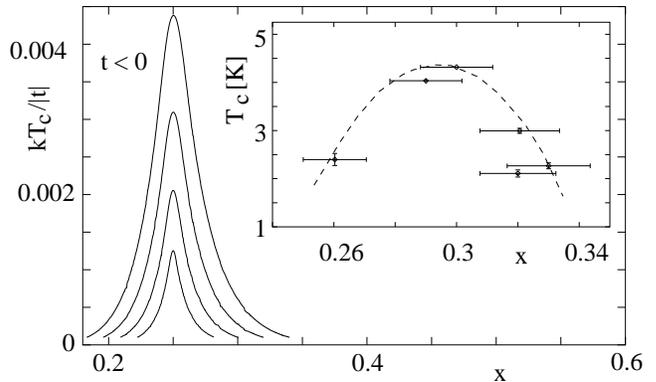}
\caption{$T_{c}$ as a function of doping for $t<0$. The curves from the
bottom to the top correspond to $J/|t|=$0.35, 0.4, 0.45, 0.5. For
comparison, the insert shows experimental data for $\mathrm{Na}_{x}\mathrm{%
CoO_{2}}\cdot y\mathrm{H_{2}0}$ taken from Ref. \protect\cite{tomek}.}
\label{result}
\end{figure}

The RVB mean field treatment of the SC phase of the Hamiltonian (\ref{4.5})
is now based on the following assumptions:

i) the global \ SU(2)$\times$U(1) symmetry is spontaneously broken
by a local order parameter down to SU(2). The SU(2) symmetry is the
exact symmetry of the SC phase; ii) the dynamics of the SC phase is
governed by the BCS--type dynamics of the valence bond \emph{hole}
SU(2) singlet pairs and is determined by the
linearized hole--hole interaction $J\sum_{ij}\left( \overrightarrow{M%
}_{i}\overrightarrow{M}_{j}-\frac{1}{4}\widehat{\delta }_{i}\widehat{\delta }%
_{j}\right) $ as well as by the hopping term. The hole spin singlets
interact with the quasiclassical spinon background $J\sum_{ij}\left(
\overrightarrow{S}_{i}\overrightarrow{S}_{j}-\frac{1}{4}\right) $ via the
induced moment-moment interaction $J\sum_{ij}\overrightarrow{S}_{i}%
\overrightarrow{M}_{j}.$ This can be treated within the MF
approximation as well. However, for the cobaltates we ignore, at
first, quantum fluctuation effects of the spinon field in comparison
with the ones originated by the $\Psi -$field.~\cite{sf} In
contrast, for the cuprates the $\langle
\overrightarrow{S}_{i}\overrightarrow{M}_{j}\rangle $ correlation
functions seem to be of crucial importance and should therefore be
treated beyond a MF approximation. This is confirmed by the
observation of antiferromagnetic ordering associated with the
superconducting vortex cores;~\cite{lake} iii) the constrained RVB
hole-singlet annihilation operator takes the form $%
\ B_{ij}=\Psi _{i\uparrow }\Psi _{j\downarrow }-\Psi _{i\downarrow }\Psi
_{j\uparrow }$, $B_{ii}=0.$ In the SC phase the $U(1)$ global \ symmetry $%
\Psi _{j\sigma }\rightarrow e^{i\theta }\Psi _{j\sigma }$ is spontaneously
broken by the local SU(2) invariant order parameter $\Delta
_{ij}=\left\langle B_{ij}\right\rangle .$

The $t$--$J$ Hamiltonian function in Eq.(\ref{4.3})
now reduces to
\begin{eqnarray}
H_{t-J}^{SC} &=&t\sum_{ij\sigma }
\bar{\Psi} _{i\sigma }\ \Psi _{j\sigma }+ h.c.
+ \frac{JNZ|\Delta |^{2}}{4}-N_{e}\mu  \nonumber \\
&&+\frac{J}{2}\sum_{ij}\left( \Psi _{i\uparrow }\Psi _{j\downarrow
}-\Psi _{i\downarrow }\Psi _{j\uparrow }\right) \overline{\Delta
}_{ij}+h.c. \label{5.1}
\end{eqnarray}%
where the chemical potential $\mu $ has been introduced to control
the number of electrons, $\widehat{N}_{e}= N-\sum_{\sigma i}\Psi
_{i\sigma }^{+}\ \Psi _{i\sigma }.$ This Hamiltonian continues to be
invariant under the $SU(2)$ action induced by (\ref{4.1}).

Despite its similar appearance with the standard MF RVB
result\cite{bza}, eq.(\ref{5.1}) has a different content. The most
important difference is related to the presence of constraints
imposed by the $\delta$--functions in Eq. (\ref{4.2}). However, even
if these functions are neglected, the  Hamiltonian (\ref{5.1}) deals
directly with the dopant--particle operators ${\Psi
_{i\sigma}}^{\prime}$s rather than with the electron operators
${c_{i\sigma}}^{\prime}$s. In terms of ${\Psi
_{i\sigma}}^{\prime}$s, $\widehat{N}_{e}$ has a different
representation, and a different equation for the chemical potential
follows from that. As a consequence, within our approach a nonzero
$\Delta $ directly implies superconductivity in contrast to the
original BZA approach, where at half-filling $\Delta ^{BZA}$ is
non-zero, but the state is insulating.

Due to the fact that we are directly dealing with the dopant
particles $T_c$ vanishes for $\delta=0$. Moreover, the average of
the kinetic term in $H^{SC}_{t-J} \sim t\delta$. The increase of
$\delta$ reflects itself in the gain of kinetic energy which will
eventually be of the order of $J$. When this occurs the singlet
pairs tend to break up and the SC phase disappears. Strictly
speaking $T_c$ is non-zero for any non-vanishing $\delta$ (see Fig.
1). This is an artifact of the applied MF approximation since it
neglects a possible onset of magnetic ordering.\cite{lee2} In the
standard RVB decoupling scheme the $J$--term vanishes above the RVB
transition temperature. Therefore, the resulting phase diagram for
the square lattice usually does not include the AF phase for the
half-filled case. In the present approach the J-term is expressed as
a sum of spinon, dopant and spinon-dopant terms (see Eq.
(\ref{4.5})). The introduced above RVB state naturally disappears at
half-filling and the Hamiltonian (\ref{4.5}) is reduced to the
Heisenberg one. Therefore, for $\delta=0$, the system becomes
insulating and the ground state energy can only be lowered by the
onset of the magnetic phase. In particular, for the square lattice
one expects the onset of the long range order antiferromagnetic
order. In order to determine the actual boundary of the magnetic
phase, one should consider it together with the RVB phase, since
these phases compete with each other.

A few remarks are in order. Since the RVB singlets are doped
hole--hole pairs  the present MF favors larger hole doping in
contrast to
the BZA scheme. Thus the present approach must be more reliable for the $t$--%
$J$ interaction on a triangular lattice.

It can be shown (see Appendix) that the equations for the order parameter
and the chemical
potential that follow from the partition function representation (\ref{4.2}%
) with the MF BCS Hamiltonian (\ref{5.1}) are invariant under the change $%
t\rightarrow -t,\ $ $\delta\rightarrow 1-\delta,$ $\ \mu \rightarrow
-\mu .$\ Thus the NDO constraint imposes
within the MF BCS approximation a symmetry restriction on a
possible structure of the phase diagram. Namely, the phase
diagrams $T_{c}(\delta )$ at $t>0$ and $t<0$
must be located symmetrically with respect to the point
$\delta =\frac{1}{2}$.
 Any renormalization of the type $t\rightarrow \delta t$,
frequently used in order to implement the constraint of no double occupancy in the
MF BCS scheme,
evidently spoils this symmetry.

Finally, Eq. (\ref{5.1}) corresponds to hole doping.\ However,
as already mentioned earlier on, the $\mathrm{CoO}_{2}$'s
are more likely electron doped compounds. In order to deal with
this case, within the representation (\ref{5.1}), we make a
canonical transformation $\Psi _{\sigma }\rightarrow \Psi
_{-\sigma }^{+}$ and keep the NDO constraint as
before. In its new form the operator $\Psi _{\sigma }^{+} $
creates a dopant electron. The phase diagram $T_{c}\times \delta $
which follows from our new ``dual'' RVB scheme is shown in Fig. 1,
for hole doping. If we replace $\delta \rightarrow x,t\rightarrow
-t$ we reproduce the main figure for the electron doping case. Our
results for this case are shown in Fig. 2. In the insert of this
figure we reproduce the experimental data from Schaak
{\it et al.}\cite{tomek} for comparison.

This phase diagram is evaluated directly from Eq. (\ref{5.1})
considering a triangular lattice of $\mathrm{CoO_{2}^{\prime }}$s.
The $d+id$ symmetry of the MF OP predicted earlier in
Refs. \onlinecite{baskaran,shastry} and \onlinecite{hon} is employed throughout the
calculations. Other symmetries can be tested if necessary using the
same scheme. The representation (\ref{4.2}) with the Hamiltonian
function given by Eq. (\ref{5.1}) incorporates the NDO constraint
rigorously and tells us that at most one spinful
fermion can live on a given lattice site. Technically, the problem
reduces to a computation of the fermionic determinant in the
presence of the constraints imposed by the $\delta$-functions.

The fermionic determinant arises upon
integrating a bilinear form in the exponential over the complex spinors $%
\Sigma _{\overrightarrow{k},\varpi _{n}}\equiv \left( \overline{\Psi }%
_{\uparrow \overrightarrow{k},\varpi _{n}},\Psi _{\downarrow -%
\overrightarrow{k},-\varpi _{n}}\right)$. Here $\varpi _{n}=\frac{\pi }{%
\beta }(2n+1)$ stands for the Matsubara fermionic frequency and vector $%
\overrightarrow{k}$ $\in $ $BZ$. Had there been no $\delta$-functions
in~(\ref{4.2}) the amplitudes
$\Sigma _{\overrightarrow{k},\varpi _{n}> 0}$ and
$\Sigma _{\overrightarrow{k},\varpi _{n}< 0}$ would have been completely
independent and contributed equally to the partition function.
In the presence of the $\delta$-functions, however,
those amplitudes are no longer independent.

The $\delta$--functions result in some interference between
these amplitudes reducing the total contribution to the
partition function. In order to estimate this reduction at
the mean--field level we use the following trick.
We multiply the piece of the free energy that comes from
the evaluation of the determinant at the absence of the
constraints by a coefficient $\kappa<1$. Then, requiring that
resulting equations for the order parameter and the chemical
potential be invariant under the change $t \to -t$,
$\delta \to 1-\delta$, and $\mu \to -\mu$ gives $\kappa=1/2$.

Although this approximation cannot be justified rigorously, it goes
beyond the one based on the renormalization of the hopping term in
the form, $t\to \delta t$, which is frequently used to partly take
into account the restriction of no double occupancy. In particular,
our approximation does not spoil the already mentioned symmetry of
the MF phase diagram under the changes $t\rightarrow -t,\ \delta
\rightarrow 1-\delta $ dictated by the NDO constraint
(see Fig. 1 and the Appendix). However, a more detailed
analysis must take into account a rigorous treatment of the
delta--function contribution.

Our results for $\mathrm{Na}_{x}\mathrm{CoO_{2}}\cdot
y\mathrm{H_{2}0}$ are very suggestive since  the
experimentally observed dome structure of the phase diagram is
reproduced by theoretical calculations within a RVB framework. The
obtained widths for the dome are also of the same magnitude as given
by experiments,~\cite{tomek,foo} although our doping values are
somewhat shifted toward the origin.
However, recent experimental results\cite{milne} indicate that the
actual hole concentration in the cobalt planes may differ from
that estimated solely on the basis of the ${\mathrm Na}$ content
and the optimal doping can be shifted from the value
reported in Ref.\onlinecite{tomek}.
The precise value of $J$ for
this compound is
still unknown. However for $t=-0.1$eV and $J/|t|$ ranging from $0.35$ to $%
0.5 $ as depicted in Figure 2, max $T_{c}$ varies roughly from
$1K$ to $4K$. Our mean field results are, therefore, in good
agreement with the existing experimental data.

 The obtained phase diagram is asymmetric with respect
to the change $t \rightarrow -t$ (electron and hole doping). The  $t
\rightarrow -t$ asymmetry has also been obtained in Ref.
\onlinecite{lee1} within a MF slave--boson Hamiltonian. In both MF
approaches this  asymmetry is an obvious consequence of the
free--particle dispersion relation on the triangular lattice. In our
case, this asymmetry concerns only the different values of the
optimal doping in electron-- and hole--doped systems. In Ref.
\onlinecite{lee1}, it is associated predominantly with the different
width of the SC region in the $T_c\times\delta$ phase diagram for
different doping regimes. Additionally, we have obtained much larger
value of the optimal doping than that reported in Ref.
\onlinecite{lee1}. Note, that the maximal value of $T_c$ obtained in
Ref. \onlinecite{lee1} for the case of electron doping is close to
that obtained for the hole doping.

According to our knowledge, there is however no experimental
verification concerning the explicit form of this asymmetry in
contrast to the electron--hole asymmetry observed in cuprate
superconductors. The $t$--$J$ Hamiltonians (\ref{1.1}) on a square
lattice with double/zero occupancy for hole/electron doping are
unitary equivalent. Accordingly, on a square lattice there is no
asymmetry with respect to the change $t \rightarrow -t$. In fact,
the doping asymmetry in the high--$T_c$ cuprates has quite a
different origin. A possible resolution of this puzzle has recently
been provided within a two--species $t$-$J$ model in Ref.
\onlinecite{bask_new}.

The doping dependence of the superfluid stiffness $D_s$ is an important ingredient
of the standard RVB theory as discussed in Ref. \onlinecite{pvt}. In particular, small values
of $D_s$ for $\delta \rightarrow 0$ determine the superconducting transition temperature.
This quantity  can be obtained with the help of the linear response theory from
the relation between the current and the transverse gauge field.
The response kernel consists of paramagnetic and diamagnetic parts. In the superconducting state
(or more precisely for $\Delta \neq 0 $) the paramagnetic contribution vanishes
for $T \rightarrow 0$.\cite{scal1} It has been shown that the diamagnetic part imposes also
the upper bound on $D_s$.\cite{param3}
In the case of the hypercubic lattice with the nearest--neighbor hopping, $D_s/\pi$
is bounded by the absolute value of the kinetic energy on a bond, whereas
for a more general dispersion relation $\varepsilon(\vec k)$ the kinetic energy should be replaced by \cite{param3}
$$
K(T)= \frac{1}{N} \sum_{\vec k \in BZ} n(\vec k) {\rm Tr}\left[ m^{-1}(\vec k)\right],
$$
where $m^{-1}_{ab}(k)= \partial^{2} \varepsilon(\vec k) / \partial  k_a  \partial k_b $.
In order to estimate the magnitude of the superfluid stiffness in our approach, we have calculated $T=0$
limit of this quantity.
Fig 3. shows the doping dependences of $K(0)$ and $\Delta(0)$.
In the standard RVB approach the NDO constraint $t \rightarrow \delta t $ renormalizes the superfluid
stiffness since it modifies the kinetic energy.
Consequently, close to half--filling $D_s$ vanishes despite the finite value of $\Delta$.\cite{pvt}
In this regime $T_c \sim D_s \sim \delta$. In our approach the superfluid stiffness
vanishes for the half--filling as well.
Since the kinetic energy term in Eq. (\ref{4.5})
contains only the dopant particles the vanishing
of the superfluid stiffness for $\delta \rightarrow 0$ is an intrinsic feature of our approach
and occurs independently of the applied approximations.
\begin{figure}[tbp]
\vspace*{3mm} \includegraphics[width=8.5cm]{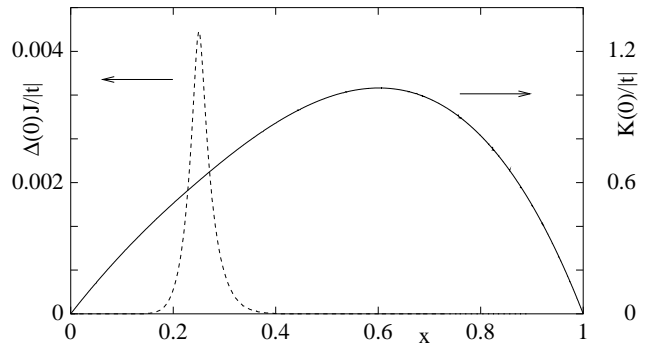}
\caption{Doping dependence of $K(T)$ and $\Delta(T)$ obtained for the triangular lattice with
$J/|t|=$0.5 and $T \rightarrow 0$.}
\label{result}
\end{figure}

We end this section discussing briefly how one can control
the BCS mean--field
decoupling (\ref{5.1}) within the su$(2|1)$ supersymmetric representation of the
Hubbard operators. This can be done by means of a large--$N$ expansion based
on a generalization of the SU(2) globally invariant $t$--$J$
Hamiltonian (\ref{1.2}) in terms of the
symplectic group Sp(2N) of 2N$\times$ 2N unitary matrices (note that
Sp(2)$\cong$ SU(2))~\cite{read,vojta,pepin}:
\begin{eqnarray}
H^{SU(2)}_{t-J}\to H^{Sp(2N)}_{t-J}=-t\sum_{ij}X^{\sigma 0}_iX^{0\sigma}_j
+h.c.  && \nonumber \\ + \frac{J}{N}\sum_{ij}\epsilon_{\sigma\sigma'}\epsilon_{\eta\eta'}
X^{\sigma\eta'}X^{\sigma'\eta}-\mu\sum_iX^{\sigma\sigma}_i, &&
\label{24.1}
\end{eqnarray}
where the summation over the Sp(2N) indexes $\sigma,\eta=\pm 1,\pm 2,...,\pm N$
is assumed and the Sp(2N) antisymmetric tensor $\epsilon_{\sigma\sigma'}$=
sgn$(\sigma)\delta_{\sigma,-\sigma'}$.
The local constraint of the $t$--$J$ model can now be taken in the form
$\sum_{\sigma}X^{\sigma\sigma}_i=N$ at every lattice site.

The exchange term in (\ref{24.1}) can be rewritten in terms of the
Sp(2N) invariant valence - bond operators
$\tilde{B}^{+}_{ij}:=\sum_{\sigma\sigma'}\epsilon_{\sigma\sigma'}
X^{\sigma 0}_iX^{\sigma' 0}_j$ in the form
$$ -\frac{J}{N}\sum_{ij}\tilde{B}^{+}_{ij}\tilde{B}_{ij}$$
and may be decoupled by the link field
$$\tilde{\Delta}_{ij}:=\frac{1}{N}<\sum_{\sigma\sigma'}\epsilon_{\sigma\sigma'}
X^{\sigma 0}_iX^{\sigma' 0}_j>.$$
At $N=\infty$ this decoupling becomes exact.

The su$(2|1)$
supersymmetric representation of the Hubbard operators
can be extended
to the case of the su$(2N|1)$ superalgebra. Since Sp(2N)$\subset$ SU(2N) one
can then employ the su$(2N|1)$ coherent states and the corresponding
path integral to treat the t-J Hamiltonian (\ref{24.1}).
In this way we will eventually arrive at an Sp(2N) globally invariant
generalization of the representation~(\ref{4.5}), with $\vec S$ and $\vec M$ operators
being now replaced by the corresponding Sp(2N) generators.
More details will be given elsewhere.

\section{Conclusion}

To conclude we developed a RVB mean field theory which takes a
direct account of the dopant carriers. These dopant particles are
represented  by appropriate fermion fields which carry both spin and
charge and transform themselves as true SU(2) spinors. The resulting
theory is written in a very convenient form since we are able in
this way to consider the both doping dependence of the critical
temperature as well as the kinetic energy effects which eventually
destroy the superconductivity at larger dopings. By making a more
extensive use of Hubbard operators we go beyond the conventional
slave-boson approximation and take sufficient care of all symmetry
properties of the Hamiltonian model. Since we apply a mean--field
decoupling, there is always a gap in the energy spectrum of the
dopant particles whenever $\Delta \neq 0 $. Consequently, the
superconducting transition temperature $kT_c$ and the energy gap $J
\Delta $ are of the same order. In the slave--boson RVB formulation
of Ref. \onlinecite{lee1}, for low doping  $T_c$ corresponds to the
condensation of holons. Therefore, in that approach $T_c$ is
decoupled from the value of $\Delta$. Such a decoupling may also
occur within our approach, e.g., when spinons are considered beyond
the mean--field level. One can see from Eq. (\ref{6.1}) that the
lowest--order spinon--holon coupling takes on the form $
\frac{J\Delta}{2}\sum_{ij}\xi_i\xi_j(\bar z_j-\bar z_i) + h.c. $.
The resulting density of states for holons may be finite also for
$\Delta \neq 0 $.\cite{bfm}

We initially applied this new RVB scheme to describe the
superconducting properties observed in the cobaltates. We succeeded
in getting qualitative good agreement with experiment. The dome
structure of the phase diagram $T_{c}\times\delta$ is well
reproduced within a RVB framework~\cite{liu}. This is
achieved without any symmetry violation of the $t$-$J$ model for the
whole doping regime.

While preparing this version of our work we came
across another RVB formulation in terms of dopant carriers
\cite{Ribeiro}. Those authors use an extended $t$-$J$ model with $t$, $t'$
and $t''$ hopping parameters. In their scheme however those parameters
are renormalized by interactions and this procedure automatically
violates the underlying symmetries of the original $t$-$J$ model.

As discussed in the Appendix, the MF phase diagram $T_c(\delta)$ for
the $t$-$J$ model on the square lattice without frustration results
in a max $T_c$ located at $\delta=1/2.$ Although incorporating the
next-nearest-neighbor (NNN) interaction in the kinetic term slightly
shifts the diagram toward the origin, it cannot account for the
experimentally observed curve for the cuprates. It's clear that
frustration is an important ingredient for the success of our RVB
method. However, apart from that there is yet another important
feature with needs to be taken into consideration to properly deal
with the cuprates case. In the cuprates there are strong
antiferromagnetic correlations which manifest themselves even inside
the superconducting vortex cores. As a result the
$<\overrightarrow{S}_{i} \overrightarrow{M}_{j}>$ correlations,
which seem unimportant for the cobaltates, may also play an
important role in the cuprates.\cite{marcin} This will produce
strong phase fluctuations which, most likely, need to be taken into
account beyond mean field approximation. This work is in progress
and will be presented elsewhere.

As a final concluding remark let us say a few words about the
ideology of the present paper. The basic idea is to use Hubbard
operators, instead of the standard fermion operators accompanied
with the nonholonomic constraint of no double occupancy. This
enables us to impose the NDO constraint locally at each lattice
site. This constraint results in strong electron correlation effects
which are  believed to be essential ingredients for dopped Mott
insulators. Since the Hubbard operators appear as Gutzwiller
projected (GP) electron operators on the states with no double
occupancy, it is in principle reasonable to work directly with the
GP operators and wave functions. In this way Paramekanti, Randeria
and Trivedi recently studied the Hubbard model making use of
parameters relevant for the cuprates, in the framework of the
variational Monter Carlo GP d-wave state~\cite{randeria}. They
showed that the strong electron correlations imposed by the
Gutzwiller projection destroy the off diagonal long range order as
$\delta \to 0$ qualitatively tracking the observed nonmonotonic
$T_c(\delta)$. Basically the same result follows from the Plain
Vanilla version of RVB, where the Gutzwiller projection is treated
within the mean-field representation.

In our approach we also treat the NDO constraint at the mean-field
level. However, we go a step further since we make explicit use of
the algebraic relations between the Hubbard operators, namely those
of the su$(2|1)$ superalgebra. This adds some extra information
which is encoded in the superalgebra commutation relations. In
particular the classical phase space realization (the coherent-state
representation) of su$(2|1)$ provides us with the complex canonical
coordinates $(z,\, \xi)$ which eventually appear as the basic
spinon-fermion fields in the path integral effective action
(\ref{2.1}). Dopant quasiparticle amplitudes (\ref{4.0}) are
constructed out of these fields appearing in our theory as emergent
phenomena. We arrive naturally in this way at the RVB theory for
dopant carriers.

\begin{acknowledgments}
One of us (E.K.) wants to acknowledge the hospitality of the ICCMP's staff
and the financial support received from CAPES - Brazil.
\end{acknowledgments}

\section{Appendix}
In this Appendix we show that within the BCS MF approximation (\ref{5.1})
the equations for the order parameter and the chemical potential
are invariant under the change $%
t\rightarrow -t,\ $ $\delta\rightarrow 1-\delta,$ $\ \mu \rightarrow
-\mu,$ provided
the NDO constraint is rigorously taken into account.

First, we integrate out the fields
$\Psi _{\uparrow i},\overline{\Psi }_{\uparrow i}$
in Eq.~(\ref{4.2}) with the Hamiltonian function given by (\ref{5.1}), which
results in
the effective action (\ref{3.4}) with the classical Hamiltonian function
now being,
\begin{eqnarray}
&&H^{cl}_{SC}=-t\sum_{ij}(\xi _{i}\overline{\xi }_{j}\langle
z_{i}|z_{j}\rangle +h.c.) -\mu
\left(N-\sum_i\bar\xi_i\xi_i\right) \nonumber \\
&&+\frac{J\Delta}{2}\sum_{ij}\left(\xi_i\xi_j
\frac{\bar z_j-\bar z_i}{\sqrt{(1+|z_i|^2)(1+|z_j|^2)}} +h.c\right).
\label{6.1}
\end{eqnarray}
Here $z_i(t)$ and $\xi_i(t)$ are dynamical fields.
This representation rigorously incorporates the constraint of no double occupancy.
Because of the rather complicated form of the action, we are in general unable
to write out explicitly a quantum counterpart of Hamiltonian~(\ref{6.1})
as a function of the su(2) spin generators
and spinless U(1) fermionic operators. However, in the SC phase we
get $z_i(t)=z_i$,
which means that quantum fluctuations of the background spinon fields are
ignored. In that case only the fermionic kinetic term is left in the
action (\ref{3.4}), and the quantum Hamiltonian can be easily identified,
\begin{eqnarray}
&&H_{SC}=-t\sum_{ij}(f_{i}f^{\dagger}_{j}\langle z_{i}|z_{j}\rangle
+h.c.) -\mu
\left(N-\sum_if^{\dagger}_if_i\right) \nonumber \\
&&+\frac{J\Delta}{2}\sum_{ij}\left(f_if_j \frac{\bar z_j-\bar
z_i}{\sqrt{(1+|z_i|^2)(1+|z_j|^2)}} +h.c\right). \label{6.2}
\end{eqnarray}
The $f_i$'s stand for the on-site spinless fermionic operators, with
$\{f_i,f^{\dag}_j\}=\delta_{ij}$ that correspond to the classical
Grassmann amplitudes, $f^{cl}=:\xi$, which give
$\{\xi_i,\overline{\xi }_{j}\}=0$. The dynamical spinon field
$z_i(t)$ looses its time-dependence and turns itself therefore into
a sort of external classical $c$-valued spinon field.

Next, we evaluate the on-site free energy function,
\begin{equation}
F/N=-\frac{1}{N}Tr\,e^{-\beta H_{SC}}, \label{6.3}\end{equation} where
the symbol $Tr$ is used to indicate the summation over the fermionic
degrees of freedom as well as the complex $c$-valued spinon fields:
\begin{equation}
Tr\,(\cdots):=\int D\mu_{su(2)}(\bar z,z)\,tr_{f,f^{\dag}}\,(\cdots)
\label{6.4}\end{equation} The $z$-integral in (\ref{6.4}) appears as
an ordinary multiple integral. In this way the order parameter and
chemical potential are determined by the conditions $\partial
F/\partial\Delta=0, \,\partial F/\partial\mu=\delta-1$ which
explicitly give
\begin{equation}
\left\langle \frac{J}{2}\sum_{ij}f_if_j
\frac{\bar z_j-\bar z_i}{\sqrt{(1+|z_i|^2)(1+|z_j|^2)}} +h.c\right\rangle=0,
\label{6.5}\end{equation}
and
\begin{equation}
\left\langle \frac{1}{N}\sum_{i}f^{\dag}_if_i\right\rangle=\delta,
\label{6.6}\end{equation} respectively. Here
$\langle(\cdots)\rangle:=Tr\,(\cdots)e^{-\beta H_{SC}}/ Tr\,e^{-\beta
H_{SC}}$. It can be checked straightforwardly that eqs.~(\ref{6.5}) and
(\ref{6.6}) are invariant under the change $t\to -t,\, \mu \to
-\mu,\, \delta\to 1-\delta.$ To see this one should simultaneously
make the canonical transformation, $f_i\to f^{\dag}_i$, and change
the integration variables, $\,z_i\to -\bar z_i.$ Accordingly, the
phase diagrams $T_c(\delta)$ at $t>0$ and $t<0$ are located
symmetrically with respect to the point $\delta=1/2.$

Explicitly, the equations for the order parameter and chemical
potential read
\begin{eqnarray}
\frac{1}{N}\sum_{\vec k\in BZ}
\frac{\tanh(\frac{E_{\vec k}\beta}{2})}{E_{\vec k}}
\mid \beta_{\vec k} \mid^2=\frac{Z}{J},
\label{eq:6.7}\end{eqnarray}
\begin{eqnarray}
\frac{1}{2N}\sum_{\vec k\in BZ}\frac{\tanh(\frac{E_{\vec k}\beta}{2})}{E_{\vec k}}
(t_{\vec k}-\mu) =\delta-1/2,
\label{eq:6.8} \end{eqnarray}
where
\begin{equation}
E^2_{\vec k}=(t_{\vec k}-\mu)^2 +J^2\Delta^2 \mid \beta_{\vec k}
\mid^2, \label{eq:6.9}\end{equation} and $t_{\vec k}=-2t\gamma_{\vec
k},\, \gamma_{\vec k}=\sum_{\vec n} \cos \vec k{\vec n}. $ In the
case of the 2D square lattice $\gamma_{\vec k}=\cos k_x+\cos k_y$,
whereas $\gamma_{\vec k}=\cos k_x + 2 \cos(k_x/2) \cos(k_y
\sqrt{3}/2)$ for the 2D triangular lattice. For the $d_{x^2-y^2}$
pairing on the square lattice the phase factor reads $\beta_{\vec
k}= \cos k_x-\cos k_y$. For the triangular lattice we assume a
$d_1+id_2$ symmetry of the order parameter.\cite{shastry,baskaran}
Then,
\begin{equation}
\beta_{\vec k}=\cos k_x -\cos \frac{k_x}{2} \cos \frac{k_y \sqrt{3}}{2} +i \sqrt{3}
\sin \frac{k_x}{2} \sin \frac{k_y \sqrt{3}}{2}.
\end{equation}
The equations (\ref{eq:6.7}) and (\ref{eq:6.8}) are clearly seen to be invariant under the change $t\to -t,\,
\mu\to -\mu,\, \delta\to 1-\delta,$ which results in the phase diagram
depicted on Fig.1.

Note that the $t$-$J$ Hamiltonian on a square lattice with the nearest-neighbor
(NN) interaction
is invariant under the change, $t\to -t$. This is because this change
amounts to a certain unitary transformation of the lattice electron operators.
It then follows that the above two phase diagrams merge in this case
into one, located at $\delta=1/2.$ Incorporating frustration
(e.g., by taking into account the NNN interaction in the t-dependent term)
destroys this symmetry and results in splitting of this diagram again into
two located symmetrically with respect to the point $\delta=1/2.$
However for the generic values of the t-J parameters that splitting
is rather small
and cannot account for an experimentally observed phase diagram for the
cuprates.

If we ignored completely the NDO constraint taking into account the
modes $\Sigma _{\overrightarrow{k},\varpi _{n}>0}$ and $\Sigma
_{\overrightarrow{k},\varpi _{n}<0}$ on equal grounds, we would get
(on a square lattice) a diagram with max $T_c$ located at
$\delta=1$. This is markedly different from the NDO constraint-free
BZA result, where max $T_c$ occurs at $\delta=0$, which bears out
that our theory is in a sense dual to the original BZA approach.

The conventional BZA MF theory formulated in terms of the lattice
electron spin singlets with the renormalization $t\to \delta t$
being implemented to partly incorporate the NDO constraint, however
fails to maintain the symmetry of the phase diagram dictated by this
constraint, and results in the same observation: max $T_c$ takes
place again at $\delta=0$, as in the constraint-free BZA theory. To
see this consider the BZA MF Hamiltonian,~\cite{bza}
\begin{eqnarray}
&&H_{t-J}^{BZA} =-t\delta\sum_{ij\sigma}c_{i\sigma}^{+}c_{j\sigma}+ h.c
-\mu\sum_{i\sigma}c^{\dag}_{i\sigma}c_{i\sigma}
 \nonumber \\
&&+\frac{J\Delta}{2}\sum_{ij}\left(c_{i\uparrow}c_{j\downarrow}
-c_{i\downarrow}c_{j\uparrow} +h.c. \right) + \frac{JNZ|\Delta |^{2}}{4}. \label{6.10}
\end{eqnarray}
One obtains the following system of equations to determine the order
parameter and chemical potential:
\begin{eqnarray}
\frac{1}{N}\sum_{\vec k}\frac{\tanh(\frac{E_{\vec k}\beta}{2})}{E_{\vec k}}
\gamma_{\vec k}^2=\frac{Z}{2J},
\label{eq:6.11}\end{eqnarray}
\begin{eqnarray}
\frac{1}{N}\sum_{\vec k}\frac{\tanh(\frac{E_{\vec k}\beta}{2})}{E_{\vec k}}
(t_{\vec k}-\mu) =\delta,
\label{eq:6.12} \end{eqnarray}
where
\begin{equation}
E^2_{\vec k}=(t_{\vec k}-\mu)^2 +J^2\Delta^2\gamma_{\vec k}^2,
\label{eq:6.13}\end{equation} and $t_{\vec k}=-2t\delta\gamma_{\vec
k},\, \gamma_{\vec k}=\sum_{\vec n} \cos \vec k{\vec n}. $ The
ensuing phase diagram $T^{BZA}_c(\delta)$ is invariant under the
change $\delta\to -\delta$ so that max $T_c$ always occurs at
$\delta =0$.


\begin{thebibliography}{99}
\bibitem{takada} K. Takada, H. Sakurai, E. Takayama--Muromachi, F. Izumi,
R. A. Dilanian, and T. Sasaki, Nature {\bf 422}, 53 (2003).


\bibitem{tomek} R.E. Schaak, T. Klimczuk, M. L. Foo, and R. J. Cava,
 Nature {\bf 424}, 527 (2003).

\bibitem{foo} 
M. L. Foo, Y. Wang, S. Watauchi, H. W. Zandbergen, T. He, R. J. Cava, and N. P. Ong,
Phys. Rev. Lett. {\bf 92}, 247001 (2004)

\bibitem{baskaran} G. Baskaran, Phys. Rev. Lett. \textbf{91}, 097003 (2003).

\bibitem{shastry} B. Kumar and B. S. Shastry, 
Phys. Rev. B \textbf{68}, 104508 (2003).

\bibitem{lee1} Q. -H. Wang, D. -H. Lee, and P. A. Lee, 
Phys. Rev. B \textbf{69}, 092504 (2004).

\bibitem{lee2} O. I. Montrunich and P. A. Lee, 
Phys. Rev. B {\bf 70}, 024514 (2004).

\bibitem{baskaran1} G. Baskaran, cond--mat/0306569.

\bibitem{bza} G. Baskaran, Z. Zou, and P. W. Anderson, Solid State Commun.
\textbf{63}, 973 (1987).

\bibitem{ubbens} G. Kotliar and J. Liu, Phys. Rev. B \textbf{38}, 5142
(1988); M. U. Ubbens and P. A. Lee, Phys. Rev. B \textbf{49}, 6853
(1994).

\bibitem{spalek} K. A. Chao, J. Spa{\l}ek, and A. M. Ole{\'s}, Phys. Rev. B
\textbf{18}, 3453 (1978).

\bibitem{h} J. Hubbard, Proc. R. Soc. London, Ser. A \textbf{285}, 542
(1965).

\bibitem{w} P. B. Wiegmann, Phys. Rev. Lett. \textbf{60}, 821 (1989).

\bibitem{solomon}   A. Montorsi, M. Rasetti, and A. Solomon, Phys. Rev. Lett.
{\bf 59}, 2243 (1987).
\bibitem{pepin}  P. Coleman, C. P\'epin, and J. Hopkinson, Phys. Rev. {\bf B63},
140441(R) (2001).
\bibitem{coleman}  P. Coleman, C. P\'epin, cond-mat/0110099.


\bibitem{k2} E.A. Kochetov and M. Mierzejewski, Phys. Rev. B \textbf{61},
1580 (2000).

\bibitem{berezin}  F. A. Berezin, {\it Introduction to Superanalysis}
(Reidel, Dordrecht, 1987).

\bibitem{gauge} S. Feng, J. Qin, and T. Ma, J. Phys.: Condens. Matter {\bf 16}
343 (2004); S. Feng, Phys. Rev. B {\bf 68}, 184501 (2003);
 S. Feng, T. Ma, and J. Qin, Mod. Phys. Lett. B{\bf 17}, 361 (2003).


\bibitem{image} By a classical image of an operator algebra we
mean its realization in terms of the Poisson brackets by functions
on a classical phase space manifold.


\bibitem{rob} S. Robaszkiewich, R. Micnas, and J. Ranninger, Phys. Rev. B
\textbf{36}, 180 (1987).

\bibitem{pvt} P.W. Anderson, P.A. Lee, M. Randeria, T.M. Rice, N.
Trivedi and F.C. Zhang, J. Phys. Condens. Matter, \textbf{16}, R755
(2004).

\bibitem{sf} The spin fluctations are partly absorbed into
the definition~(\ref{4.0}).


\bibitem{lake}
B. Lake, H.M. R{\o}nnow, N.B. Christensen, K. Lefmann, G. Aeppli, D.F. McMorrow,
 N. Mangkorntong, M. Nohara, H. Takagi, T.E. Mason, P. Vorderwisch, P. Smeibidl,
Nature, \textbf{415}, 299 (2002).

\bibitem{hon} C. Honerkamp, Phys. Rev. B \textbf{68}, 104510 (2003).

\bibitem{milne} C. J. Milne, D. N. Argyriou, A. Chemseddine,
 N. Aliouane, J. Veira, S. Landsgesell, and D. Alber,
 cond-mat/0401273.

\bibitem{bask_new} G. Baskaran, cond--mat/0505509.

\bibitem{scal1} D. J. Scalapino. S. R. White, and S. Zhang, Phys. Rev. B {\bf 47}, 7995 (1993).

\bibitem{param3} A. Paramekanti, N. Trivedi, and M. Randeria, Phys. Rev. B {\bf 57}, 11639 (1998).
\bibitem{read}  N. Read and S. Sachdev, Phys. Rev. Lett. {\bf 66}, 1773 (1991).

\bibitem{vojta}  M. Vojta, Y. Zhang, and S. Sachdev, Phys. Rev. {\bf B62}
6721 (2000).

\bibitem{bfm}  A. Ferraz, E. Kochetov, M. Mierzejewski - to be published.

\bibitem{liu} After the completion of this work we have seen a paper
by Bin Liu, Ying Liang, and Shipping Feng, cond-mat/0405168, where
the observed dome structure of the superconducting transition
temperature versus doping phase diagram has also been obtained
theoretically though within quite a different framework.

\bibitem{Ribeiro} T. C. Ribeiro and X.-G. Wen,
Phys. Rev. Lett. {\bf 95}, 057001 (2005)


\bibitem{marcin} Our approach properly accounts for the
competition between superconductivity and the magnetic order. If the
SU(2) symmetry is broken $\langle \overrightarrow{S}_{i}\rangle \neq
0 $. Then, the MF hamiltonian (\ref{5.1}) should be extended by the
following term : $\sum_i \overrightarrow{h}_{i}
\overrightarrow{M}_{i}$, where
$\overrightarrow{h}_{i}=-J\sum_{\left<j\right>_i}
\overrightarrow{\langle S \rangle}_{i}$ and the summation is carried out over the
neighboring sites. The magnetic order occurs in the hamiltonian in
an analogous way as the magnetic field introduced by the Zeeman
term. This interaction, when sufficiently strong, destroys
superconductivity. Site dependence of the effective field
$\overrightarrow{h}_{i}$ depends on the lattice geometry. In the
case of antiferromagnetic order on a square lattice
$\overrightarrow{h}_{i}$ represents a stagger magnetic field.


\bibitem{randeria} A. Paramekanti, M. Randeria, and N. Trivedi, Phys. Rev. Lett.
{\bf 87}, 217002 (2001); Phys. Rev. B \textbf{70}, 054504 (2004).


\end{thebibliography}
\end{document}